\pdfoutput=1

\documentclass[12pt,preprint]{aastex}

\usepackage{amsmath,amssymb,graphicx}
\usepackage{epsfig}
\usepackage{amssymb}
\usepackage{natbib}
\usepackage{aas_macros}
\usepackage[section] {placeins}
\usepackage{subfigure}
\usepackage{float}

\def\msun{{\rm\,M_\odot}}

\def\msun{{\rm\,M_\odot}} 

\def\zsun{{\rm\,Z_\odot}}

\newcommand{\lya}{Ly$\alpha$ }

\def\h2{${\rm\,H_2}$}

\makeatletter 

\def\msun{{\rm\,M_\odot}}

\def\vol#1  {{{#1}{\rm,}\ }}
\def\lya{{\rm Ly}\alpha}

\newcount\refno
\newcount\rfno
\def\eq{$^{\the\refno\ }$\advance\refno by 1}
\def\ad{\advance\rfno by 1}

\def\clock{\count0=\time \divide\count0 by 60
     \count1=\count0 \multiply\count1 by -60 \advance\count1 by \time
     \number\count0:\ifnum\count1<10{0\number\count1}\else\number\count1\fi}

\def\myputfigure#1#2#3#4#5%
{\hskip0.03\textwidth\vskip#5pt
\makebox[0pt]{\hskip#2in
\includegraphics[width=#3\textwidth]{#1}}\vskip#4pt\hfill}

\lefthead{Cen} 
\righthead{Reionization}

\begin{document}

\title{The State of the Universe at $z\sim 6$}

\author{
Renyue Cen$^{1}$
} 
 
\footnotetext[1]{Princeton University Observatory, Princeton, NJ 08544;
 cen@astro.princeton.edu}

\begin{abstract} 

In the context stellar reionization in the standard cold dark matter model, 
we analyze observations at $z\sim 6$ and are able to draw three significant conclusions 
with respect to star formation and the state of the intergalactic medium (IGM) at $z\sim 6$.
(1) An initial stellar mass function (IMF) more efficient, by a factor of $10-20$, 
in producing ionizing photons than the standard Salpeter IMF
is required at $z\sim 6$.
This may be achieved by having either (A) a metal-enriched IMF with 
 a lower mass
cutoff of $\ge 30\msun$ or (B) $2-4\%$ of stellar mass being Population III
massive metal-free stars at $z\sim 6$.
While there is no compelling physical reason or observational evidence to support (A), 
(B) could be fulfilled plausibly by continued existence of some pockets of uncontaminated, metal-free gas
for star formation. 
(2) The volume-weighted neutral fraction of the IGM of
$\langle f_{\rm HI}\rangle_V\sim 10^{-4}$ at $z=5.8$ inferred from
the SDSS observations of QSO absorption spectra
provides enough information to ascertain that reionization is basically complete 
with at most $\sim 0.1-1\%$ of IGM that is un-ionized at $z=5.8$.
(3) Barring some extreme evolution of the IMF, the neutral fraction of the IGM is expected to rise quickly
toward high redshift from the point of HII bubble percolation, 
with the mean neutral fraction of the IGM 
expected to reach $6-12\%$ at $z=6.5$, $13-27\%$ at $z=7.7$ and $22-38\%$ at $z=8.8$.
\end{abstract}

 
\section{Introduction}

How the universe becomes transparent at $z\sim 5.8$ is debated \citep[][]{2006Fan,2007Becker}.
Whether reionization is complete by $z=5-6$
has been questioned \citep[][]{2009bMesinger}. 
What kind of stars reionizes the universe at $z\sim 6$ remains less than certain.
We examine in greater detail this endgame
to assess how reionization
process may have proceeded approaching $z\sim 5.8$,
how complete reionization is at $z\sim 5.8$
and what role Population III (Pop III) stars may have played in
the final reionization phase at $z\sim 6$,
in the context of stellar reionization 
in the standard cold dark matter model \citep[][]{2010Komatsu}.
We are also motivated by the exciting possibility of 
being able to statistically measure the neutral fraction 
of the IGM at redshift above six in the coming years,
as a variety of techniques are applied to larger samples that will become available.
Those include methods based on (1) QSO Stromgren sphere measures
\citep[e.g.,][]{2004Wyithe,2004Mesinger},
(2) measurements of damping wings of high redshift gamma-ray bursts (GRB) 
\citep[e.g.,][]{2006Totani},
(3) statistical analyses of high redshift Lyman alpha emitters 
from a variety of surveys
\citep[e.g.,][]{2004Malhotra,2007Ouchi, 2008Ouchi,
2007Nilsson,
2007Cuby,
2007Stark,
2008Willis,
2008McMahon,
2009Hibon}.
In addition, polarization measurements of the cosmic microwave 
background (CMB) fluctuations by Planck satellite and others may provide
some useful constraints \citep[e.g.,][]{2003Kaplinghat}.
Finally, the James Webb Space Telescope (JWST) will likely be able to detect
the bulk of dwarf galaxies of halo mass $\sim 10^9\msun$ that are believed to be 
primarily responsible for cosmological reionization
at $z\sim 6$ \citep[e.g.,][]{2004Stiavelli}, 
especially if a significant fraction of stars in them are active Pop III stars.

\section{Evolution of the Intergalactic Medium Toward $z\sim 6$}

We use a semi-numerical method \citep{2003Cen} to 
explore the parameter space and compute
the coupled thermal and reionization history with star formation of the universe.
The reader is referred to \S4 of \citet{2003Cen} for details.
For a simple understanding 
the essential physics pertaining to reionization may be encapsulated into 
a single parameter, $\eta$, defined as 
\begin{equation}
\label{eq:sum}
\eta(z) \equiv {c_* f_{\rm esc} R_h(z) \epsilon_{\rm UV}(z) m_{\rm p} c^2\over \alpha(T) C(z) n_0 (1+z)^3 h\nu_0},
\end{equation}
\noindent 
where $z$ is redshift,
$c_*$ the star formation efficiency (i.e., the ratio of 
the total amount of stars formed over the product of the halo mass
and the cosmic baryon to total mass ratio),
$f_{\rm esc}$ the ionizing photon escape fraction,
$R_h(z)$ the total baryonic mass accretion rate of halos above the filter mass
(i.e., those that are able to accrete gas) over the total baryonic mass in the universe,
$\epsilon_{\rm UV}(z)$ the ionizing photon production efficiency, 
defined to be the total emitted energy above hydrogen Lyman limit over
the total rest mass energy of forming stars,
$m_{\rm p}$ proton mass, $c$ speed of light,
$\alpha(T)$ the case-B recombination coefficient, 
$C(z)$ the clumping factor of the recombining IGM,
$n_0$ the mean hydrogen number density at $z=0$
and $h\nu_0$ hydrogen ionization potential.
The numerator on the right hand side of Equation \ref{eq:sum}
 is the rate of ionizing photons per baryon pumped into the IGM from stars,
whereas the denominator is the destruction rate of Lyman limit photons per baryon due to case-B
recombination.
If $\eta < 1$, the universe is opaque.
When $\eta > 1$ is sustained, the universe becomes fully reionized
and a UV radiation background is built up with time with its amplitude
determined by the balance between UV emissivity, recombination and universal expansion.
If $\eta$ goes above unity at an earlier epoch and subsequently drops below unity,
a double reionization would occur \citet{2003Cen}.
Present calculations are done with the following updates of input physics.
\begin{itemize}
\item We adopt the standard WMAP7-normalized \citep[][]{2010Komatsu}
parameters for the cosmological constant dominated, flat 
cold dark matter model:
$\Omega_M=0.28$, $\Omega_b=0.046$, $\Omega_{\Lambda}=0.72$, $\sigma_8=0.81$,
$H_0=100 h {\rm km s}^{-1} {\rm Mpc}^{-1} = 70 {\rm km} s^{-1} {\rm Mpc}^{-1}$ 
and $n=0.96$.
\item We replace the standard Press-Schechter formalism 
of spherical collapse model with the more accurate ellipsoidal collapse model  \citep[][]{2002Sheth} 
to compute the halo formation rate $R_h$. 
\item Latest ultra-high resolution ($0.1$pc) radiation hydrodynamic simulations
indicate that $c_* f_{\rm esc} \sim 0.02-0.03$ 
for atomic cooling halos
and drops about two order of magnitude for minihalos,
with $f_{\rm esc} \sim 40-80\%$ \citep[][]{2009Wise}.
Note that $c_* f_{\rm esc}$ and $\epsilon_{\rm UV}(z)$ are degenerate.
Therefore, we adopt, conservatively, $c_* f_{\rm esc}=0.03$ for the calculations presented here, 
which enables a firm conclusion with respect to a required high value for $\epsilon_{\rm UV}(z)$, 
as will be clear later.
\item We allow for an evolving IMF with redshift, parameterized by an evolving 
ionizing photon production efficiency, 
$\epsilon_{\rm UV}(z)=\epsilon_{\rm UV,6} ({1+z\over 7})^\gamma$,
where $\epsilon_{\rm UV,6}$ is $\epsilon_{\rm UV}(z)$ at $z=6$.
\item The clumping factor, $C(z)$, of the recombining IGM
at $z\sim 6$ may be lower than previous estimates.
We adopt the suggested range $C_6=3-6$ for the clumping factor at $z=6$ 
based on recent calculations \citep[][]{2009Pawlik}.
\end{itemize}

In our semi-numerical method, the evolution of the clumping factor of the IGM is
determined by one parameter, $C_{h}$, that takes into account the contribution of collapsed gas to
the overall clumping factor: $C(z) = \phi_h(z) C_{h} + [1-\phi_h(z)]$,
where $\phi_h(z)$ is the fraction of mass in halos above the filter mass
\citep[][]{2000bGnedin} that is followed self-consistently; 
we adjust $C_{h}$ along with the other free parameter,
$\epsilon_{\rm UV}(z)$, until we obtain simultaneously a desired clumping factor at 
$z=6$, $C_6$, and that the universe completes reionization at exactly $z=5.8$.
We also examine a case where reionization ends at $z=6.8$.

\begin{figure}[h]
\centering
\vskip -1.7in
\epsfig{file=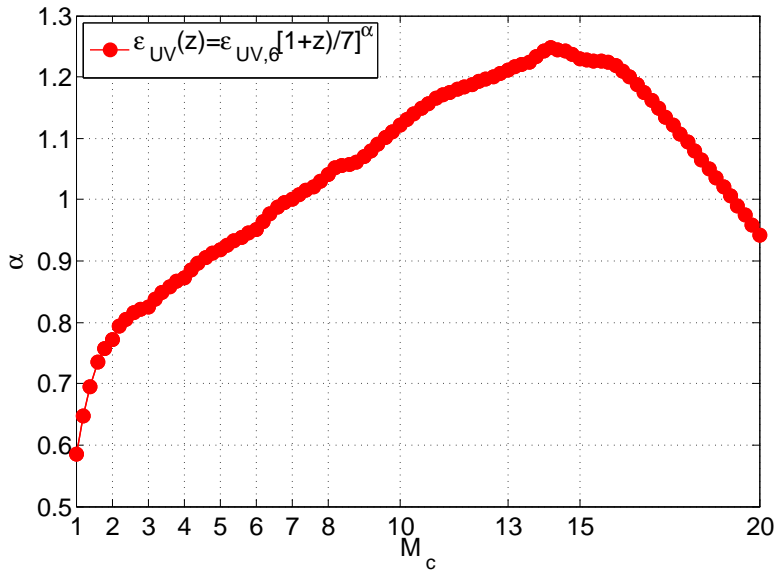,angle=0,width=4.5in}
\vskip -1.5in
\caption{
The mean slope of the expected evolution 
$\epsilon_{\rm UV}(z)=\epsilon_{\rm UV,6} ({1+z\over 7})^\gamma$
from $z=9$ to $z=6$ for an evolving IMF with a Salpeter slope
and a varying lower mass cutoff at $z=6$ shown at the x-axis.
}
\label{fig:alpha}
\end{figure}

Perhaps the most uncertain of the input physics on the list is $\epsilon_{\rm UV}(z)$,
which we now elaborate on.
For a fiducial, non-evolving IMF case $\epsilon_{\rm UV}(z)=\epsilon_{\rm UV,6}$.
For an evolving IMF, we take cue from recent development in the field of star formation 
at high redshift, in particular, on CMB-regulated star formation physical process
\citep[e.g.,][]{2005Larson,2007Tumlinson, 2009Smith, 2010Bailin, 2010Schneider}.
Following \citet[][]{2007Tumlinson},
the CMB-regulated Bonner-Ebert mass 
of a collapsing cloud evolves as
$M_{\rm BE} = 3.2 [(1+z)/7]^{1.7} \msun$.
Specifying lower mass cutoff ($M_{\rm c}$) of a Salpeter IMF at $z=6$
and assuming that evolves as $M_{\rm c}[(1+z)/7]^{1.7}$,
and using the Padova $0.02\zsun$ track \citep[][]{1999Leitherer} to obtain 
$\epsilon_{\rm UV,6}$ and $\epsilon_{\rm UV}(z=9)$,
we compute $\gamma$ as a function of $M_{\rm c}$, shown in Figure~\ref{fig:alpha}.
Depending on the exact value of $M_{\rm c}$,
$\gamma$ ranges from $0.6$ to $1.25$ for $M_{\rm c}=1-20\msun$.
While it is uncertain,
we identify $M_{\rm BE} = 3.2 \msun$ at $z=6$ 
with $M_{\rm c}$, giving rise to $\gamma=0.84$.
In our subsequent analyses, we treat $\gamma=0$ and $\gamma=0.84$
as two limiting cases for the evolution of IMF.

\begin{figure}[h]
\centering
\vskip -1.4in
\epsfig{file=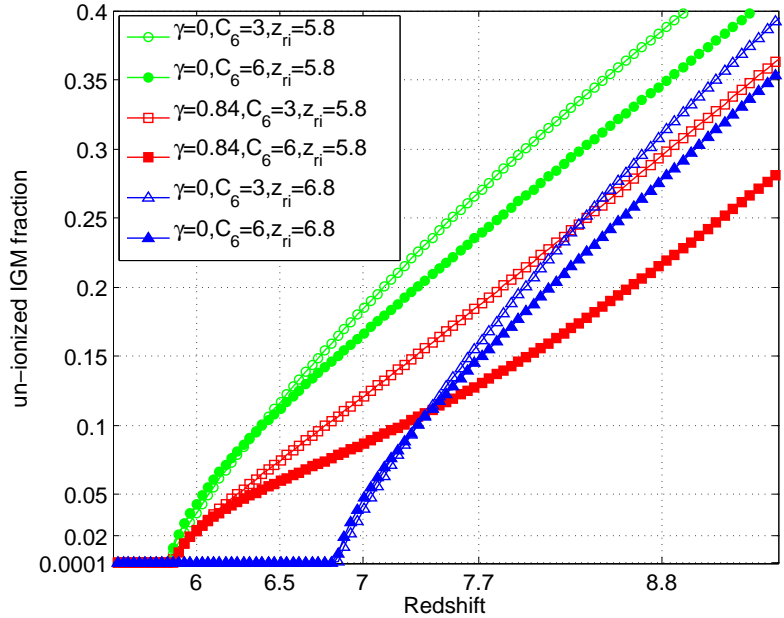,angle=0,width=4.5in}
\vskip -1.2in
\caption{
The evolution of un-ionized fraction of the IGM, $x$, in six
different reionization models with specified $\epsilon_{\rm UV}(z)$, $C_6$ and $z_{\rm ri}$.
}
\label{fig:y}
\end{figure}

With $c_*f_{\rm esc}=0.03$, $\gamma$, $C_6$
and the completion redshift of reionization $z_{ri}$ being fixed,
we can find a unique pair of values for $C_h$ and $\epsilon_{\rm UV,6}$.
Figure~\ref{fig:y} shows the evolutionary histories of the fraction 
of the un-ionized IGM, $x$, for six models. 
A feature common in all six models is that
$x$ rapidly rises toward higher redshift from $z_{ri}$.
Analysis of the SDSS observations of QSO absorption spectra
suggests a transition to a (volume-weighted) neutral fraction
$\langle f_{\rm HI}\rangle_V\ge 10^{-3}$ at $z\sim 6.2$ from 
$\langle f_{\rm HI}\rangle_V\sim 10^{-4}$ at $z=5.8$ \citep[][]{2006Fan}.
As we will show below, the observed $\langle f_{\rm HI}\rangle_V\sim 10^{-4}$ at $z=5.8$ 
indicates that the reionization is largely complete by $z=5.8$.
Thus, our models suggest that
$x$ is expected to reach $6-12\%$ at $z=6.5$, $13-27\%$ at $z=7.7$ and $22-38\%$ at $z=8.8$.

It is useful to have some simple physical understanding of the results.
Star formation and reionization is somewhat self-regulated
in that a higher star formation rate ionizes and heats up a larger fraction of 
the IGM that would tend to suppress gas accretion for further star formation,
whereas cooling processes induce more star formation \citep{2003Cen}.
As the response time scale for this self-regulation is on the order of
the halo dynamic time that is roughly 10\% of the Hubble time,
so if there is a protracted period during reionization,
this argument suggests that it may only take place 
at a neutral fraction level of $x\ge 10\%$ so as 
to allow star formation rate to be able to dynamically respond to reionization 
induced heating within a halo dynamical time.
Once $x$ has dropped significantly below $10\%$,
the final stage of reionization should be prompt, 
which is greatly conspired by the rapid increase of $\eta(z)$ toward the end of reionization
(Equation \ref{eq:sum}).
It can be shown that $R_h\propto \exp(-\delta^2_c/2\sigma^2_M) (1+z)$, leading to
$\eta \propto \exp(-\delta^2_c/2\sigma^2_M) (1+z)^{\gamma-2} C^{-1}(z)$,
where $\sigma_M$ is the density variance on the mass scale of $M\sim 10^9\msun$ 
that can accrete photoheated gas and form stars \citep[][]{2000bGnedin}.
By the end of reionization, about $1\%$ of total mass turns out to have 
collapsed in these halos; in other words, the (star-forming) halo collapse rate 
is on the exponential rise when the universe becomes fully ionized.
Since the evolution of $C$ is much weaker than exponential \citep[][]{2009Pawlik},
$\eta(z)$ likely surpasses unity at $z_{ri}$
in an ``exponential" fashion from below.
As a result, it takes significantly less than $x$ times Hubble time to reionize the
last small $x$ fraction of neutral IGM.
These considerations are 
consistent with the rapid final reionization phase seen in Figure~\ref{fig:y}.

The SDSS observations strongly suggest $z_{ri}=5.8$ \citep[][]{2006Fan},
after which the ionization state of the IGM is primarily determined, 
on the ionizing photon sink side, by LLS.
We show here, from a somewhat different angle, but in agreement with the conclusion of \citet[][]{2006Fan},
that reionization is largely complete by $z=5.8$ 
\citep[c.f.,][]{2009bMesinger}. 
The comoving mean free path (mfp) of Lyman limit photons, $\lambda$, 
may be written as
\begin{equation}
\label{eq:local}
\lambda^{-1}=\lambda_{\rm LL}^{-1} + \lambda_{\rm Lya}^{-1}+\lambda_{\rm neu}^{-1} + \lambda_{\rm other}^{-1}, 
\end{equation}
\noindent 
where $\lambda_{\rm LL}$, $\lambda_{\rm Lya}$ and $\lambda_{\rm neu}$
are comoving mfp due to LLS, $\lya$ forest and un-ionized neutral IGM, respectively;
$\lambda_{\rm other}$ is due to possible other sinks.
Physically, LLS are in less ionized, overdense regions
within the reionized portion of the universe that are individually opaque to Lyman limit photons;
$\lya$ forest is dominated by low density regions 
within the reionized portion of the IGM 
that individually are only partially opaque to Lyman limit photons;
the un-ionized neutral IGM is the portion of the IGM that has not been engulfed
by the reionization front.
We conservatively assume $\lambda_{\rm other}=\infty$.
Current large-scale cosmological reionization 
simulations do not provide sufficiently accurate results to 
constrain $\lambda_{\rm LL}$ due to lack of adequate resolution. 
An extrapolation \citep[][]{2006Gnedin} 
of observations at lower redshift $z=0.4-4.7$ 
\citep[][]{1994Storrie-Lombardi} gives $\lambda_{\rm LL}\sim 22-48$ comoving Mpc/h 
at $z\sim 5.8$.
We use $\lambda_{\rm LL}=35$ comoving Mpc/h in our calculations.
The following three equations are used to compute the neutral fraction
of a region, $x_\delta$, at overdensity $\delta\equiv \rho_b/\langle\rho_b\rangle$,
when the region is substantially ionized (i.e., $x_\delta \ll 1$):
\begin{equation}
\label{eq:local}
x_\delta J_\nu \langle\sigma_H\rangle = \delta \alpha(T) n_0 (1+z)^3,
\end{equation}
\noindent 
where $J_\nu$ is the ionizing photon radiation intensity in units of cm$^{-2}$~sec$^{-1}$
and $\langle\sigma_H\rangle = 2.6\times 10^{-18}$cm$^2$ is the spectrum-averaged 
photoionization cross section
for a low-Z IMF ionizing spectrum at high-z.
\begin{equation}
\label{eq:global}
\Psi = C \alpha(T) n_0 (1+z)^3,
\end{equation}
\noindent 
where $\Psi$ is mean ionizing photon emissivity per baryon.
\begin{equation}
\label{eq:emis}
J_\nu = \lambda \Psi n_0 (1+z)^2.
\end{equation}
\noindent 
\begin{figure}[h]
\centering
\vskip -1.0in
\epsfig{file=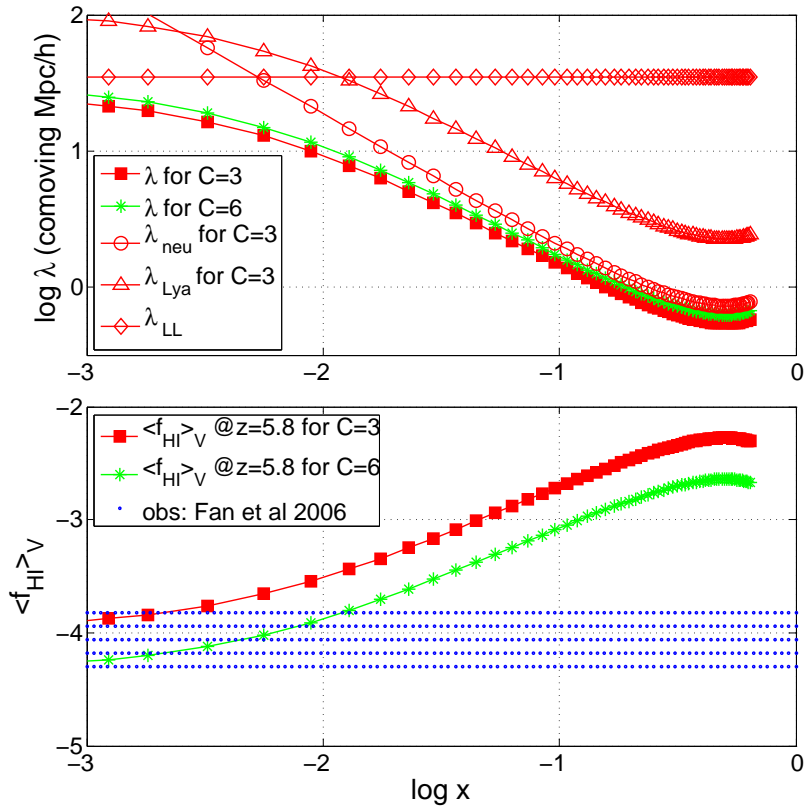,angle=0,width=4.5in}
\vskip -0.9in
\caption{
Top panel shows the total comoving mean free path (mfp) $\lambda$ 
for two cases with $C=3$ (solid squares)
and $C=6$ (stars) as well as 
$\lambda_{\rm LL}$ (open diamonds), 
$\lambda_{\rm Lya}$ (open triangles) 
and
$\lambda_{\rm neu}$ (open circles) for the case with $C=3$,
as a function of $x$ at $z=5.8$.
Bottom panel shows the volume-weighted neutral fractions of the IGM,
$\langle f_{\rm HI}\rangle_V$,
as a function of $x$ for the two cases with $C=3$ (solid squares)
and $C=6$ (stars).
Also shown as the shaded region is the total range of 
$\langle f_{\rm HI}\rangle_V$ at $z=5.8$ based on the SDSS QSO sample \citep[][]{2006Fan}.
}
\label{fig:mfp}
\end{figure}
\noindent 
Equations \ref{eq:local},\ref{eq:global},\ref{eq:emis}, respectively,
reflect the local ionization balance (between photoionization and recombination) (3),
global ionization balance (between mean emissivity and recombination) (4)
and relationship between mean emissivity, ionizing photon intensity and mfp (5).
Combining (\ref{eq:local},\ref{eq:global},\ref{eq:emis}) we obtain 
\begin{equation}
\label{eq:xdelta}
x_\delta = {\delta \over \lambda C \langle\sigma_H\rangle n_0 (1+z)^3}.
\end{equation}
\noindent 
Thus, knowing $C$ and $\lambda$ allows one to compute $x_\delta$,
which, when combined with the probability distribution function of $\delta$, PDF$(\delta)$,
can be used to compute the volume-weighted neutral fraction, $\langle f_{\rm HI}\rangle_V$:
\begin{equation}
\label{eq:pdf}
\langle f_{\rm HI}\rangle_V = \int_0^\infty {\rm PDF}(\delta) x_\delta d\delta.
\end{equation}
\noindent 
We use the density distribution, PDF$(\delta)$,
from one of the radiation-hydrodynamic simulations \citep[][]{2008Trac}
where the universe completes reionization at $z\sim 6$,
to compute  $\langle f_{\rm HI}\rangle_V$ at $z=5.8$.
A resolution of comoving $65$kpc/h in the simulation 
is adequate for resolving the Jeans scale of photo-ionized gas.
The mfp due to $\lya$ forest can be computed as
\begin{equation}
\label{eq:lambdalya}
\lambda_{\rm Lya}^{-1} = \langle f_{\rm HI}\rangle_V\langle\sigma_H\rangle n_0 (1+z)^2/(1+1/e).
\end{equation}
\noindent 
We use the same simulation to also compute $\lambda_{\rm neu}$, simply by computing the average
distance that a random ray can travel before it hits an un-ionized cell.
We identify regions that have not been reionized and photon-heated with
cells in the simulation box that have neutral fraction greater than $0.99$ and temperature lower than $10^3$K;
results are insensitive to reasonable variations of the parameters:
changing $0.99$ to $0.50$ or $10^3$K to $100$K makes no visible difference in the results.
Since, when scaled to $x$, the morphology of reionization does not vary strongly 
\citep[e.g.,][]{2004Furlanetto},
we use $\lambda_{\rm neu}(z)$ computed as a function of redshift from the simulation
as $\lambda_{\rm neu}(x)$ as a function of $x$ at $z=5.8$.
The detailed procedure to simultaneously compute $\lambda$ and 
$\langle f_{\rm HI}\rangle_V$ is as follows.
At a given value of $x$ at $z=5.8$, we know $\lambda_{\rm neu}(x)$ from simulations.
Combining $\lambda_{\rm neu}(x)$ with an initial guess for $\lambda_{\rm Lya}(x)$ and the adopted
$\lambda_{\rm LL}$ gives $\lambda$ (Equation \ref{eq:sum}).
With $\lambda$ and an assumed $C$ we compute $\langle f_{\rm HI}\rangle_V$ 
(Equations \ref{eq:xdelta},\ref{eq:pdf}),
which in turn yields a new value for $\lambda_{\rm Lya}(x)$ (Equation \ref{eq:lambdalya}).
This procedure is iterated until we have a converged pair of
$\lambda_{\rm Lya}$ and $\langle f_{\rm HI}\rangle_V$.
The results are shown in Figure~\ref{fig:mfp},
where the top panel shows the total comoving mfp for two cases with $C=3$ and $C=6$ 
as well as various components for the case with $C=3$,
and the bottom panel shows $\langle f_{\rm HI}\rangle_V$ for 
two cases with $C=3$ and $C=6$ at $z=5.8$ as well as the observationally inferred range 
at $z=5.8$ \citep[][]{2006Fan}.
As it turns out, we see that the total $\lambda$ is primarily determined by 
$\lambda_{\rm neu}$ at $x\ge 0.01$ and 
$\lambda_{\rm LL}$ at $x\le 0.005$,
and $\lya$ forest has secondary importance at all $x$.
A comparison between the computed results and observations
indicates that the un-ionized fraction $x$ does not exceed 
$0.1-1\%$ at $z=5.8$ and reionization is complete or largely complete by $z=5.8$.
In combination with our previous finding of rapid reionization near $z_{ri}$,
it suggests that $z_{ri}=5.8$ or very near it.

\begin{figure}[h]
\centering
\vskip -0.7in
\epsfig{file=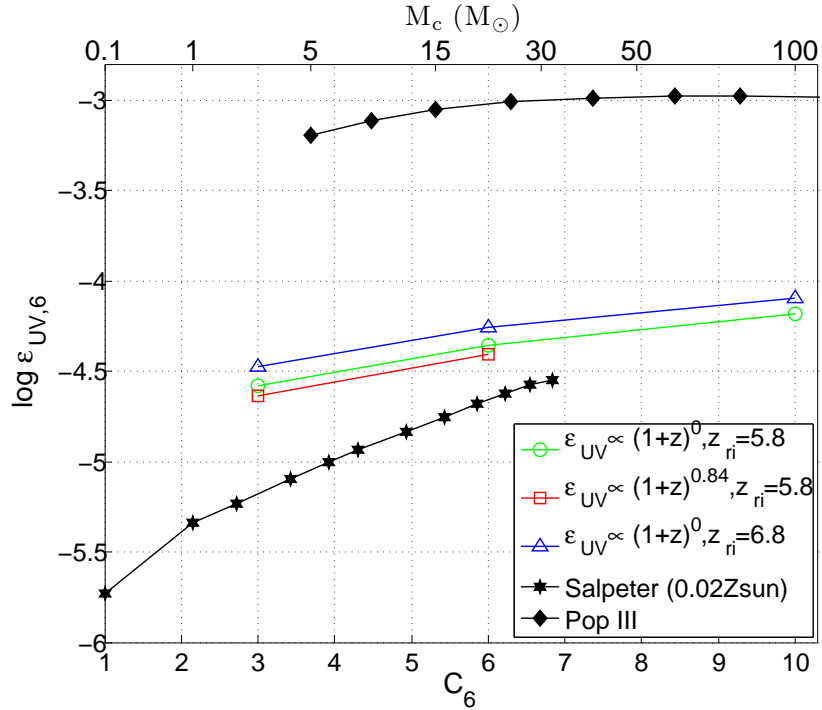,angle=0,width=4.5in}
\vskip -1.2in
\caption{
show the required ionizing photon production efficiency 
at $z=6$, 
$\epsilon_{\rm UV,6}$,
as a function of IGM clumping factor at $z=6$, $C_6$,
for several models.
Also shown as stars are the expected $\epsilon_{\rm UV}$ for 
Salpeter IMF (with $0.02\zsun$ metallicity) 
\citep[][]{1999Leitherer} 
with the lower mass cutoff $M_{\rm c}$ indicated by the top x-axis.
The diamonds are $\epsilon_{\rm UV}$ for Pop III metal-free stars, again,
with the lower mass cutoff $M_{\rm c}$ indicated by the top x-axis
\citep[][]{2002Schaerer}.
}
\label{fig:eUV}
\end{figure}

Finally, in Figure~\ref{fig:eUV} we 
show the required ionizing photon production efficiency
at $z=6$, 
$\epsilon_{\rm UV,6}$, as a function of $C_6$, for several models.
Several expected trends are noted.
First, a higher $C_6$ requires a higher $\epsilon_6$. 
Second, an earlier reionization 
requires a higher $\epsilon_6$.
Third, a rising $\epsilon_{\rm UV}$ with redshift 
lessens the required $\epsilon_6$ fractionally.
What is most striking is that stars with the standard metal-enriched IMF and $M_{\rm c}=1\msun$
fall short of providing the required ionizing photons, by a factor of $10-20$ at $z\sim 6$.
Having $M_{\rm c}\sim 5\msun$ would help reduce the deficit to a factor of $3-6$.
Only with $M_{\rm c}\sim 30\msun$ and $C_6=3$, one is barely able to meet
the requirement to reionize the universe at $z\sim 6$ by Population II stars.
But such an extreme scenario with $M_{\rm c}\ge 30\msun$ may be disfavored 
by the existence of old stars in observed high-z galaxies
\citep[e.g.,][]{2005Mobasher} that need be less massive than $\sim 10\msun$ to be long-lived.

However, we note that Pop III metal-free stars (diamonds in Figure~\ref{fig:eUV}), 
thought to be more massive than $30\msun$
\citep[e.g.,][]{2002Abel, 2002Bromm, 2008McKee},
could provide ample ionizing photons. 
Unfortunately, normally, Pop III stars would not be expected to form at $z\sim 6$, had
some earlier supernovae uniformly enriched the intergalactic medium.
With gaseous low-temperature coolants,
it is believed that the critical metallicity 
for transition from Pop III to Pop II IMF
is $Z_{\rm crit}\sim 10^{-3.5}\zsun$ \citep[e.g.,][]{2001Bromm,2003cBromm}.
If dust is formed in the Pop III.1 supernova ejecta,
\citet{2006Schneider} argue that
dust cooling may significantly lower the critical transition metallicity to as low as
$Z_{\rm crit}\sim 10^{-6}$ \citep[e.f.,][]{2010Cherchneff}.
If a fraction $10^{-4}$ of baryons forms into Pop III stars
and their supernovae uniformly enrich the IGM,
the expected metallicity of the IGM will likely
exceed $Z\sim 10^{-3.5}\zsun$ \citep[e.g.,][]{2004Fang}.
Since a fraction of $\ge 10^{-4}$ of baryons needs to form into Pop III stars to reionize the universe,
therefore, in the case of uniform IGM enrichment,
the contribution of Pop III stars to ionizing photon budget 
at $z\sim 6$ is expected to have become negligible.
In addition, a very small amount of metals ($Z\le 10^{-3}\zsun$) 
would change the internal dynamics of massive stars (core temperature, size, effective surface temperature, etc)
and render them much less efficient UV producers and 
notably different from Pop III massive stars \citep[e.g.,][]{2008Hirschi},
as already hinted in Figure~\ref{fig:eUV} between stars ($Z=0.02\zsun$) and diamonds (Pop III)
at $M_{\rm c}\sim 30\msun$.

We suggest that, if the metal enrichment process of gas, including IGM and gas in collapsed minihalos 
and other galaxies, is highly inhomogeneous,
then it is possible that a small fraction of star-forming gas may have remained 
primordial to allow for Pop III star formation at $z\sim 6$.
A significant amount of gas in the central regions of non-star-forming galaxies
\citep[e.g.,][]{2007Wyithe, 2008Cen}
as well as a fraction of IGM that has not been swept 
by galactic winds emanating from star-forming galaxies could remain uncontaminated.
Cosmological simulations at lower redshift ($z=0-6$) suggest that
metal enrichment process of the IGM is indeed extremely inhomogeneous,
leaving significant pockets of metal-free gas even at $z=0$ 
\citep[e.g.,][]{1999bCen,2001Aguirre,2006Oppenheimer,2010Cen}.
The common assumption is that 
earlier generations of stars not resolved in these simulations
would have put in a metallicity-floor in all regions.
But this needs not be the case.
Observationally, while the majority of local star formation has metallicity close to solar,
relatively low-metallicity ($1/30$ of solar) 
star formation does occur occasionally \citep[e.g.,][]{1999Izotov} 
and some of the observed local supernovae
may be pair-instability supernovae \citep[e.g.,][]{2007Smith, 2009GalYam}
that may be due to metal-free progenitors
\citep[c.f.,][]{2007Smith, 2007Langer, 2007Woosley}.
At redshift $z=2-3$ the low density $\lya$ forest, regions of density around and less than the global mean,
appears to have not been enriched to a detectable level ($\le 10^{-3.5}\zsun$)
\citep[e.g.,][]{1998Lu}.
Therefore, it seems plausible that
an increasing fraction of star-forming gas toward high redshift
may be pristine, due to a combination of inefficient and non-uniform mixing
and a decreasingly amount of metals having been injected.

From Figure~\ref{fig:eUV} we see that 
what is minimally required in order to have enough ionizing photons at $z\sim 6$
is that about 2-4\% of stars forming at $z\sim 6$
are Pop III stars and the remainder normal Pop II metal-enriched stars with a Salpeter-like IMF or other forms;
the lower mass cutoff for the latter is unconstrained but $M_{\rm c}\sim 3\msun$ or so
is perhaps physically motivated and fully in line with other evidence that hints
on an evolving IMF from redshift zero \citep[e.g.,][]{2008vanDokkum,2008Dave}.
With 2-4\% being Pop III stars, Pop III stars' contribution to ionizing photons and FUV are dominant 
over the remaining normal stars, 
which would give rise to an ``apparent", very low metallicity, top-heavy IMF for these high redshift galaxies.
Interestingly, galaxies with such required properties -
a dust-free, very low metallicity, top-heavy IMF with a very high ionizing photon escape fraction 
of $40-80\%$ - 
may have already been detected in the Hubble Ultra Deep Field (UDF) at $z\sim 7-8$
\citep[e.g.,][]{2010Bouwens}.

\section{Conclusions and Discussion}

We study the evolution of the IGM at $z\sim 5.8$ when 
universe finally turns transparent in the context of stellar 
reionization in the standard cold dark matter model. 
Under the conservative assumption, $c_* f_{\rm esc}=0.03$, 
that is based on recent calculations of ionizing photon escape fraction $f_{\rm esc}$
and star formation efficiency $c_*$, 
we find that metal-enriched Pop II stars with a normal IMF fail, 
by a factor of $10-20$, to provide 
enough ionizing photons to reionize the universe at $z\sim 6$.
Only under the scenario that the vast majority of Pop II stars are 
more massive than $\sim 30\msun$ one may be able to reionize the universe,
if the clumping factor of the recombining IGM is $\le 3$ at $z=6$.
Perhaps the observed existence of high-z galaxies at $z\sim 6$ already rules out this scenario.

Alternatively, we suggest that, if a mass fraction of $2-4\%$ of stars forming 
at $z\sim 6$ is massive Pop III metal-free stars,
enough ionizing photons can be produced.
Physically, this scenario can be plausibly accommodated by 
existence of a fraction of uncontaminated, metal-free gas at $z\sim 6$,
due to non-uniform mixing of metals in the universe.
The dominant contribution of UV light from these Pop III stars, likely being dust-free 
and expected very high ionizing 
photon escape fraction ($40-80\%$) would combine to 
make these galaxies look very blue, which would be consistent with recent observations of
$z=7-8$ galaxies in the UDF.  This hints exciting possibilities for JWST in detecting 
signatures of Pop III stars.

Based on extant observations that the volume-weighted neutral fraction 
of the IGM is $\sim 10^{-4}$ at $z\sim 5.8$,
we conclude that the reionization is basically complete by $z=5.8$
with no more than 0.1-1\% of the IGM remaining neutral.
In other words, $z_{ri}=5.8$ or very near it, in agreement with earlier, independent analyses 
\citep[][]{2006Fan}.

With the complete percolation of HII bubbles occurring at $z=5.8$, 
the mean neutral fraction of the IGM 
is expected to reach $6-12\%$ at $z=6.5$, $13-27\%$ at $z=7.7$ and $22-38\%$ at $z=8.8$.
Future measurements shall test this stellar reionization paradigm in the standard model.

Finally, we mention the possibility of probing other physics with reionization.
Significant alteration of the properties of dark matter particles from ``vanilla cold"
may have substantial impact on the reionization process.
Weakly interacting massive particle (WIMP) annihilation heating may 
affect the balance of cooling and heating processes and hence change
the primordial star formation process, if dark matter particles
are sufficiently cold to allow for very concentrated profiles 
at stellar scales \citep[e.g.,][]{2008Spolyar,2009Natarajan}.
Turning particles from ``cold" to somewhat ``warm" may also have profound effects on reionization.
Current astronomical observational constraints place a lower limit on dark matter particle mass
$m_x$ in the neighborhood of $0.5 - 1$~keV 
\citep{2000Narayanan, 2001bBarkana, 2005Viel, 2006Abazajian}.  
The smoothing scale, defined as the comoving half-wavelength of the mode for which
the linear perturbation amplitude is suppressed by 2, is $R_S=0.34
({\Omega_M\over 0.3})^{0.15}({h\over 0.7})^{1.3}({m_X\over{\rm keV}})^{-1.15} h^{-1}$ Mpc
\citep{2001Bode}.  For $m_X = (1,10)$ keV,
the mass smoothing scale is $(1.2\times 10^{10},4.3\times 10^6) \msun$, respectively. 
Naturally, the formation epoch
of Pop III star formation in a warm dark matter model of gravitino particle mass $m_x=15$ keV is found to be 
delayed relative to the case with cold dark matter by $10^8$ yr or from $z \sim 16$ to $z \sim 13$
\citep{2006OShea}.  
Given that even $m_x =15$ keV has a significant impact on Pop III star formation,
it seems very promising that observations of high-z galaxies will place constraints on the nature of
dark matter particles. 
With a dark matter particle mass of $\sim 1$ keV, minihalo formation would be largely suppressed
and give rise to a very different, ``favorable" scenario in which larger galaxies
form stars out of primordial gas that may otherwise have been contaminated by star
formation in minihalos.  
Thus, observations of reionization may provide a powerful probe of the physics of dark matter.

\acknowledgments

This work is supported in part by grants NNG06GI09G and NNX08AH31G.

\bibliographystyle{apj}
\bibliography{astro}

\begin{thebibliography}{67}
\expandafter\ifx\csname natexlab\endcsname\relax\def\natexlab#1{#1}\fi

\bibitem[{{Abazajian}(2006)}]{2006Abazajian}
{Abazajian}, K. 2006, \prd, 73, 063513

\bibitem[{{Abel} {et~al.}(2002){Abel}, {Bryan}, \& {Norman}}]{2002Abel}
{Abel}, T., {Bryan}, G.~L., \& {Norman}, M.~L. 2002, Science, 295, 93

\bibitem[{{Aguirre} {et~al.}(2001){Aguirre}, {Hernquist}, {Schaye}, {Katz},
  {Weinberg}, \& {Gardner}}]{2001Aguirre}
{Aguirre}, A., {Hernquist}, L., {Schaye}, J., {Katz}, N., {Weinberg}, D.~H., \&
  {Gardner}, J. 2001, \apj, 561, 521

\bibitem[{{Bailin} {et~al.}(2010){Bailin}, {Stinson}, {Couchman}, {Harris},
  {Wadsley}, \& {Shen}}]{2010Bailin}
{Bailin}, J., {Stinson}, G., {Couchman}, H., {Harris}, W.~E., {Wadsley}, J., \&
  {Shen}, S. 2010, \apj, 715, 194

\bibitem[{{Barkana} {et~al.}(2001){Barkana}, {Haiman}, \&
  {Ostriker}}]{2001bBarkana}
{Barkana}, R., {Haiman}, Z., \& {Ostriker}, J.~P. 2001, \apj, 558, 482

\bibitem[{{Becker} {et~al.}(2007){Becker}, {Rauch}, \& {Sargent}}]{2007Becker}
{Becker}, G.~D., {Rauch}, M., \& {Sargent}, W.~L.~W. 2007, \apj, 662, 72

\bibitem[{{Bode} {et~al.}(2001){Bode}, {Ostriker}, \& {Turok}}]{2001Bode}
{Bode}, P., {Ostriker}, J.~P., \& {Turok}, N. 2001, \apj, 556, 93

\bibitem[{{Bouwens} {et~al.}(2010){Bouwens}, {Illingworth}, {Oesch}, {Trenti},
  {Stiavelli}, {Carollo}, {Franx}, {van Dokkum}, {Labb{\'e}}, \&
  {Magee}}]{2010Bouwens}
{Bouwens}, R.~J., {Illingworth}, G.~D., {Oesch}, P.~A., {Trenti}, M.,
  {Stiavelli}, M., {Carollo}, C.~M., {Franx}, M., {van Dokkum}, P.~G.,
  {Labb{\'e}}, I., \& {Magee}, D. 2010, \apjl, 708, L69

\bibitem[{{Bromm} {et~al.}(2002){Bromm}, {Coppi}, \& {Larson}}]{2002Bromm}
{Bromm}, V., {Coppi}, P.~S., \& {Larson}, R.~B. 2002, \apj, 564, 23

\bibitem[{{Bromm} {et~al.}(2001){Bromm}, {Ferrara}, {Coppi}, \&
  {Larson}}]{2001Bromm}
{Bromm}, V., {Ferrara}, A., {Coppi}, P.~S., \& {Larson}, R.~B. 2001, \mnras,
  328, 969

\bibitem[{{Bromm} \& {Loeb}(2003)}]{2003cBromm}
{Bromm}, V. \& {Loeb}, A. 2003, \nat, 425, 812

\bibitem[{{Cen}(2003)}]{2003Cen}
{Cen}, R. 2003, \apj, 591, 12

\bibitem[{{Cen} \& {Chisari}(2010)}]{2010Cen}
{Cen}, R. \& {Chisari}, N.~E. 2010, ArXiv e-prints

\bibitem[{{Cen} \& {Ostriker}(1999)}]{1999bCen}
{Cen}, R. \& {Ostriker}, J.~P. 1999, \apjl, 519, L109

\bibitem[{{Cen} \& {Riquelme}(2008)}]{2008Cen}
{Cen}, R. \& {Riquelme}, M.~A. 2008, \apj, 674, 644

\bibitem[{{Cherchneff} \& {Dwek}(2010)}]{2010Cherchneff}
{Cherchneff}, I. \& {Dwek}, E. 2010, \apj, 713, 1

\bibitem[{{Cuby} {et~al.}(2007){Cuby}, {Hibon}, {Lidman}, {Le F{\`e}vre},
  {Gilmozzi}, {Moorwood}, \& {van der Werf}}]{2007Cuby}
{Cuby}, J., {Hibon}, P., {Lidman}, C., {Le F{\`e}vre}, O., {Gilmozzi}, R.,
  {Moorwood}, A., \& {van der Werf}, P. 2007, \aap, 461, 911

\bibitem[{{Dav{\'e}}(2008)}]{2008Dave}
{Dav{\'e}}, R. 2008, \mnras, 385, 147

\bibitem[{{Fan} {et~al.}(2006){Fan}, {Strauss}, {Becker}, {White}, {Gunn},
  {Knapp}, {Richards}, {Schneider}, {Brinkmann}, \& {Fukugita}}]{2006Fan}
{Fan}, X., {Strauss}, M.~A., {Becker}, R.~H., {White}, R.~L., {Gunn}, J.~E.,
  {Knapp}, G.~R., {Richards}, G.~T., {Schneider}, D.~P., {Brinkmann}, J., \&
  {Fukugita}, M. 2006, \aj, 132, 117

\bibitem[{{Fang} \& {Cen}(2004)}]{2004Fang}
{Fang}, T. \& {Cen}, R. 2004, \apjl, 616, L87

\bibitem[{{Furlanetto} {et~al.}(2004){Furlanetto}, {Zaldarriaga}, \&
  {Hernquist}}]{2004Furlanetto}
{Furlanetto}, S.~R., {Zaldarriaga}, M., \& {Hernquist}, L. 2004, \apj, 613, 1

\bibitem[{{Gal-Yam} {et~al.}(2009){Gal-Yam}, {Mazzali}, {Ofek}, {Nugent},
  {Kulkarni}, {Kasliwal}, {Quimby}, {Filippenko}, {Cenko}, {Chornock},
  {Waldman}, {Kasen}, {Sullivan}, {Beshore}, {Drake}, {Thomas}, {Bloom},
  {Poznanski}, {Miller}, {Foley}, {Silverman}, {Arcavi}, {Ellis}, \&
  {Deng}}]{2009GalYam}
{Gal-Yam}, A., {Mazzali}, P., {Ofek}, E.~O., {Nugent}, P.~E., {Kulkarni},
  S.~R., {Kasliwal}, M.~M., {Quimby}, R.~M., {Filippenko}, A.~V., {Cenko},
  S.~B., {Chornock}, R., {Waldman}, R., {Kasen}, D., {Sullivan}, M., {Beshore},
  E.~C., {Drake}, A.~J., {Thomas}, R.~C., {Bloom}, J.~S., {Poznanski}, D.,
  {Miller}, A.~A., {Foley}, R.~J., {Silverman}, J.~M., {Arcavi}, I., {Ellis},
  R.~S., \& {Deng}, J. 2009, \nat, 462, 624

\bibitem[{{Gnedin}(2000)}]{2000bGnedin}
{Gnedin}, N.~Y. 2000, \apj, 542, 535

\bibitem[{{Gnedin} \& {Fan}(2006)}]{2006Gnedin}
{Gnedin}, N.~Y. \& {Fan}, X. 2006, \apj, 648, 1

\bibitem[{{Hibon} {et~al.}(2009){Hibon}, {Cuby}, {Willis}, {Cl{\'e}ment},
  {Lidman}, {Arnouts}, {Kneib}, {Willott}, {Marmo}, \& {McCracken}}]{2009Hibon}
{Hibon}, P., {Cuby}, J., {Willis}, J., {Cl{\'e}ment}, B., {Lidman}, C.,
  {Arnouts}, S., {Kneib}, J., {Willott}, C.~J., {Marmo}, C., \& {McCracken}, H.
  2009, ArXiv e-prints

\bibitem[{{Hirschi} {et~al.}(2008){Hirschi}, {Chiappini}, {Meynet}, {Maeder},
  \& {Ekstr{\"o}m}}]{2008Hirschi}
{Hirschi}, R., {Chiappini}, C., {Meynet}, G., {Maeder}, A., \& {Ekstr{\"o}m},
  S. 2008, in IAU Symposium, Vol. 250, IAU Symposium, ed. {F.~Bresolin,
  P.~A.~Crowther, \& J.~Puls}, 217--230

\bibitem[{{Izotov} \& {Thuan}(1999)}]{1999Izotov}
{Izotov}, Y.~I. \& {Thuan}, T.~X. 1999, \apj, 511, 639

\bibitem[{{Kaplinghat} {et~al.}(2003){Kaplinghat}, {Chu}, {Haiman}, {Holder},
  {Knox}, \& {Skordis}}]{2003Kaplinghat}
{Kaplinghat}, M., {Chu}, M., {Haiman}, Z., {Holder}, G.~P., {Knox}, L., \&
  {Skordis}, C. 2003, \apj, 583, 24

\bibitem[{{Komatsu} {et~al.}(2010){Komatsu}, {Smith}, {Dunkley}, {Bennett},
  {Gold}, {Hinshaw}, {Jarosik}, {Larson}, {Nolta}, {Page}, {Spergel},
  {Halpern}, {Hill}, {Kogut}, {Limon}, {Meyer}, {Odegard}, {Tucker}, {Weiland},
  {Wollack}, \& {Wright}}]{2010Komatsu}
{Komatsu}, E., {Smith}, K.~M., {Dunkley}, J., {Bennett}, C.~L., {Gold}, B.,
  {Hinshaw}, G., {Jarosik}, N., {Larson}, D., {Nolta}, M.~R., {Page}, L.,
  {Spergel}, D.~N., {Halpern}, M., {Hill}, R.~S., {Kogut}, A., {Limon}, M.,
  {Meyer}, S.~S., {Odegard}, N., {Tucker}, G.~S., {Weiland}, J.~L., {Wollack},
  E., \& {Wright}, E.~L. 2010, ArXiv e-prints

\bibitem[{{Langer} {et~al.}(2007){Langer}, {Norman}, {de Koter}, {Vink},
  {Cantiello}, \& {Yoon}}]{2007Langer}
{Langer}, N., {Norman}, C.~A., {de Koter}, A., {Vink}, J.~S., {Cantiello}, M.,
  \& {Yoon}, S. 2007, \aap, 475, L19

\bibitem[{{Larson}(2005)}]{2005Larson}
{Larson}, R.~B. 2005, \mnras, 359, 211

\bibitem[{{Leitherer} {et~al.}(1999){Leitherer}, {Schaerer}, {Goldader},
  {Gonz{\'a}lez Delgado}, {Robert}, {Kune}, {de Mello}, {Devost}, \&
  {Heckman}}]{1999Leitherer}
{Leitherer}, C., {Schaerer}, D., {Goldader}, J.~D., {Gonz{\'a}lez Delgado},
  R.~M., {Robert}, C., {Kune}, D.~F., {de Mello}, D.~F., {Devost}, D., \&
  {Heckman}, T.~M. 1999, \apjs, 123, 3

\bibitem[{{Lu} {et~al.}(1998){Lu}, {Sargent}, {Barlow}, \& {Rauch}}]{1998Lu}
{Lu}, L., {Sargent}, W.~L.~W., {Barlow}, T.~A., \& {Rauch}, M. 1998, ArXiv
  Astrophysics e-prints

\bibitem[{{Malhotra} \& {Rhoads}(2004)}]{2004Malhotra}
{Malhotra}, S. \& {Rhoads}, J.~E. 2004, \apjl, 617, L5

\bibitem[{{McKee} \& {Tan}(2008)}]{2008McKee}
{McKee}, C.~F. \& {Tan}, J.~C. 2008, \apj, 681, 771

\bibitem[{{McMahon} {et~al.}(2008){McMahon}, {Parry}, {Venemans}, {King},
  {Ryan-Weber}, {Bland-Hawthorn}, \& {Horton}}]{2008McMahon}
{McMahon}, R., {Parry}, I., {Venemans}, B., {King}, D., {Ryan-Weber}, E.,
  {Bland-Hawthorn}, J., \& {Horton}, A. 2008, The Messenger, 131, 11

\bibitem[{{Mesinger}(2009)}]{2009bMesinger}
{Mesinger}, A. 2009, ArXiv e-prints

\bibitem[{{Mesinger} {et~al.}(2004){Mesinger}, {Haiman}, \&
  {Cen}}]{2004Mesinger}
{Mesinger}, A., {Haiman}, Z., \& {Cen}, R. 2004, \apj, 613, 23

\bibitem[{{Mobasher} {et~al.}(2005){Mobasher}, {Dickinson}, {Ferguson},
  {Giavalisco}, {Wiklind}, {Stark}, {Ellis}, {Fall}, {et~al.}}]{2005Mobasher}
{Mobasher}, B., {Dickinson}, M., {Ferguson}, H.~C., {Giavalisco}, M.,
  {Wiklind}, T., {Stark}, D., {Ellis}, R.~S., {Fall}, S.~M., {et~al.} 2005,
  ApJ, 635, 832

\bibitem[{{Narayanan} {et~al.}(2000){Narayanan}, {Spergel}, {Dav{\'e}}, \&
  {Ma}}]{2000Narayanan}
{Narayanan}, V.~K., {Spergel}, D.~N., {Dav{\'e}}, R., \& {Ma}, C. 2000, \apjl,
  543, L103

\bibitem[{{Natarajan} {et~al.}(2009){Natarajan}, {Tan}, \&
  {O'Shea}}]{2009Natarajan}
{Natarajan}, A., {Tan}, J.~C., \& {O'Shea}, B.~W. 2009, \apj, 692, 574

\bibitem[{{Nilsson} {et~al.}(2007){Nilsson}, {Orsi}, {Lacey}, {Baugh}, \&
  {Thommes}}]{2007Nilsson}
{Nilsson}, K.~K., {Orsi}, A., {Lacey}, C.~G., {Baugh}, C.~M., \& {Thommes}, E.
  2007, \aap, 474, 385

\bibitem[{{Oppenheimer} \& {Dav{\'e}}(2006)}]{2006Oppenheimer}
{Oppenheimer}, B.~D. \& {Dav{\'e}}, R. 2006, \mnras, 373, 1265

\bibitem[{{O'Shea} \& {Norman}(2006)}]{2006OShea}
{O'Shea}, B.~W. \& {Norman}, M.~L. 2006, \apj, 648, 31

\bibitem[{{Ouchi} {et~al.}(2008){Ouchi}, {Shimasaku}, {Akiyama}, {Simpson},
  {Saito}, {Ueda}, {Furusawa}, {Sekiguchi}, {Yamada}, {Kodama}, {Kashikawa},
  {Okamura}, {Iye}, {Takata}, {Yoshida}, \& {Yoshida}}]{2008Ouchi}
{Ouchi}, M., {Shimasaku}, K., {Akiyama}, M., {Simpson}, C., {Saito}, T.,
  {Ueda}, Y., {Furusawa}, H., {Sekiguchi}, K., {Yamada}, T., {Kodama}, T.,
  {Kashikawa}, N., {Okamura}, S., {Iye}, M., {Takata}, T., {Yoshida}, M., \&
  {Yoshida}, M. 2008, \apjs, 176, 301

\bibitem[{{Ouchi} {et~al.}(2007){Ouchi}, {Tokoku}, {Shimasaku}, \&
  {Ichikawa}}]{2007Ouchi}
{Ouchi}, M., {Tokoku}, C., {Shimasaku}, K., \& {Ichikawa}, T. 2007, in
  Astronomical Society of the Pacific Conference Series, Vol. 379, Cosmic
  Frontiers, ed. {N.~Metcalfe \& T.~Shanks}, 47--+

\bibitem[{{Pawlik} {et~al.}(2009){Pawlik}, {Schaye}, \& {van
  Scherpenzeel}}]{2009Pawlik}
{Pawlik}, A.~H., {Schaye}, J., \& {van Scherpenzeel}, E. 2009, \mnras, 394,
  1812

\bibitem[{{Schaerer}(2002)}]{2002Schaerer}
{Schaerer}, D. 2002, \aap, 382, 28

\bibitem[{{Schneider} \& {Omukai}(2010)}]{2010Schneider}
{Schneider}, R. \& {Omukai}, K. 2010, \mnras, 402, 429

\bibitem[{{Schneider} {et~al.}(2006){Schneider}, {Omukai}, {Inoue}, \&
  {Ferrara}}]{2006Schneider}
{Schneider}, R., {Omukai}, K., {Inoue}, A.~K., \& {Ferrara}, A. 2006, \mnras,
  369, 1437

\bibitem[{{Sheth} \& {Tormen}(2002)}]{2002Sheth}
{Sheth}, R.~K. \& {Tormen}, G. 2002, \mnras, 329, 61

\bibitem[{{Smith} {et~al.}(2009){Smith}, {Turk}, {Sigurdsson}, {O'Shea}, \&
  {Norman}}]{2009Smith}
{Smith}, B.~D., {Turk}, M.~J., {Sigurdsson}, S., {O'Shea}, B.~W., \& {Norman},
  M.~L. 2009, \apj, 691, 441

\bibitem[{{Smith} {et~al.}(2007){Smith}, {Li}, {Foley}, {Wheeler}, {Pooley},
  {Chornock}, {Filippenko}, {Silverman}, {Quimby}, {Bloom}, \&
  {Hansen}}]{2007Smith}
{Smith}, N., {Li}, W., {Foley}, R.~J., {Wheeler}, J.~C., {Pooley}, D.,
  {Chornock}, R., {Filippenko}, A.~V., {Silverman}, J.~M., {Quimby}, R.,
  {Bloom}, J.~S., \& {Hansen}, C. 2007, \apj, 666, 1116

\bibitem[{{Spolyar} {et~al.}(2008){Spolyar}, {Freese}, \&
  {Gondolo}}]{2008Spolyar}
{Spolyar}, D., {Freese}, K., \& {Gondolo}, P. 2008, Physical Review Letters,
  100, 051101

\bibitem[{{Stark} {et~al.}(2007){Stark}, {Ellis}, {Richard}, {Kneib}, {Smith},
  \& {Santos}}]{2007Stark}
{Stark}, D.~P., {Ellis}, R.~S., {Richard}, J., {Kneib}, J.-P., {Smith}, G.~P.,
  \& {Santos}, M.~R. 2007, \apj, 663, 10

\bibitem[{{Stiavelli} {et~al.}(2004){Stiavelli}, {Fall}, \&
  {Panagia}}]{2004Stiavelli}
{Stiavelli}, M., {Fall}, S.~M., \& {Panagia}, N. 2004, \apjl, 610, L1

\bibitem[{{Storrie-Lombardi} {et~al.}(1994){Storrie-Lombardi}, {McMahon},
  {Irwin}, \& {Hazard}}]{1994Storrie-Lombardi}
{Storrie-Lombardi}, L.~J., {McMahon}, R.~G., {Irwin}, M.~J., \& {Hazard}, C.
  1994, \apjl, 427, L13

\bibitem[{{Totani} {et~al.}(2006){Totani}, {Kawai}, {Kosugi}, {Aoki}, {Yamada},
  {Iye}, {Ohta}, \& {Hattori}}]{2006Totani}
{Totani}, T., {Kawai}, N., {Kosugi}, G., {Aoki}, K., {Yamada}, T., {Iye}, M.,
  {Ohta}, K., \& {Hattori}, T. 2006, \pasj, 58, 485

\bibitem[{{Trac} {et~al.}(2008){Trac}, {Cen}, \& {Loeb}}]{2008Trac}
{Trac}, H., {Cen}, R., \& {Loeb}, A. 2008, \apjl, 689, L81

\bibitem[{{Tumlinson}(2007)}]{2007Tumlinson}
{Tumlinson}, J. 2007, \apjl, 664, L63

\bibitem[{{van Dokkum}(2008)}]{2008vanDokkum}
{van Dokkum}, P.~G. 2008, \apj, 674, 29

\bibitem[{{Viel} {et~al.}(2005){Viel}, {Lesgourgues}, {Haehnelt}, {Matarrese},
  \& {Riotto}}]{2005Viel}
{Viel}, M., {Lesgourgues}, J., {Haehnelt}, M.~G., {Matarrese}, S., \& {Riotto},
  A. 2005, \prd, 71, 063534

\bibitem[{{Willis} {et~al.}(2008){Willis}, {Courbin}, {Kneib}, \&
  {Minniti}}]{2008Willis}
{Willis}, J.~P., {Courbin}, F., {Kneib}, J., \& {Minniti}, D. 2008, \mnras,
  384, 1039

\bibitem[{{Wise} \& {Cen}(2009)}]{2009Wise}
{Wise}, J.~H. \& {Cen}, R. 2009, \apj, 693, 984

\bibitem[{{Woosley} {et~al.}(2007){Woosley}, {Blinnikov}, \&
  {Heger}}]{2007Woosley}
{Woosley}, S.~E., {Blinnikov}, S., \& {Heger}, A. 2007, \nat, 450, 390

\bibitem[{{Wyithe} \& {Cen}(2007)}]{2007Wyithe}
{Wyithe}, J.~S.~B. \& {Cen}, R. 2007, \apj, 659, 890

\bibitem[{{Wyithe} \& {Loeb}(2004)}]{2004Wyithe}
{Wyithe}, J.~S.~B. \& {Loeb}, A. 2004, \nat, 427, 815

\end{thebibliography}

\end{document}